\documentclass[letterpaper,12pt]{article}
\usepackage{tabularx} % extra features for tabular environment
\usepackage{amsmath}  % improve math presentation
\usepackage[margin=1in,letterpaper]{geometry} % decreases margins
\usepackage{cite} % takes care of citations
\usepackage[final]{hyperref} % adds hyper links inside the generated pdf file
\usepackage{authblk}
\usepackage[pdftex]{graphicx}
\hypersetup{
	colorlinks=true,       % false: boxed links; true: colored links
	linkcolor=blue,        % color of internal links
	citecolor=blue,        % color of links to bibliography
	filecolor=magenta,     % color of file links
	urlcolor=blue         
}

\newcommand{\Tabref}[1]{Table\,\ref{#1}}
\newcommand{\Figref}[1]{Fig.\,\ref{#1}}
\newcommand{\Equref}[1]{Eq.\,\ref{#1}}
\begin{document}

\title{Study of the $h\gamma Z$ coupling at the ILC}
\author{Yumi Aoki$^{1}$, Keisuke Fujii$^{2}$, Sunghoon Jung$^{3}$, Junghwan Lee$^{3}$, Junping Tian$^{4}$, Hiroshi Yokoya$^{5}$
\\
on behalf of the ILD concept group}
\affil{SOKENDAI$^{1}$, KEK$^{2}$, Seoul National University$^{3}$,University of Tokyo$^{4}$, KIAS$^{5}$}

\maketitle

\begin{abstract}
  We study the $h \gamma Z$ coupling, which is a loop induced coupling in the Standard Model (SM), to probe new physics. In a global fit based on the SM Effective Field Theory, measurement of the SM $h \gamma Z$ coupling can provide a very useful constraint, in particular for the precise determination of $hZZ$ and $hWW$ couplings. At the International Linear Collider (ILC), there are two direct ways to study the $h \gamma Z$ coupling: one is to measure the branching ratio of the $h \to \gamma Z$ decay and the other to measure the cross section for the $e^+e^- \to h \gamma$ process. 
We have performed a full simulation study of the $e^+e^- \to h \gamma$ process at the 250 GeV ILC, assuming 2 ab$^{-1}$ data collected by the International Large Detector (ILD). 
The expected 1$\sigma$ bound on the effective $h\gamma Z$ coupling ($\zeta_{AZ}$) 
combining measurements of the cross section for $e^+e^- \to h \gamma$ followed by $h \to b \bar{b}$ and the $h \to \gamma Z$ branching ratio is $-0.0015<\zeta_{AZ}<0.0015$. 
The expected significance for the signal cross section in the fully hadronic $h \to WW^*$ channel is 0.09 $\sigma$ for beam polarizations of $P(e^-,e^+)=(-80\%,+30\%)$.
\footnote{
Talk presented at the International Workshop on Future Linear Colliders (LCWS2019), Sendai, Japan, 28 October-1 November, 2019. C19-10-28.}

\end{abstract}

\section{Introduction}
\begin{figure}[ht] 
        
        \centering \includegraphics[width=0.4\columnwidth]{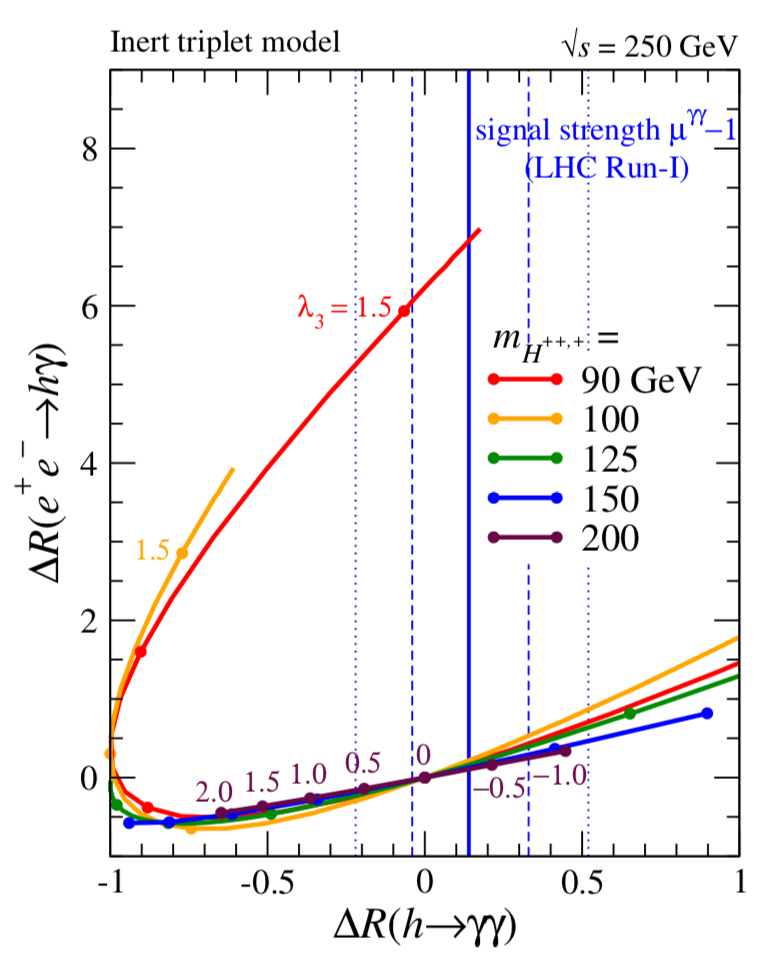}
        \caption{
                \label{fig:1} 
               The relative deviations from the Standard Model for the $e^+ e^- \to h \gamma$ cross section and the $h \xrightarrow{} \gamma\gamma$ decay branching ratio~\cite{Aus2}.}
\end{figure}

Precision study of the Higgs boson is a powerful tool to find physics beyond the standard model. The International Linear Collider (ILC)\,\cite{Aus0} is an ideal machine to carry out the precision Higgs measurements. Our motivation here is to probe new physics in $h \gamma \gamma$ and $h \gamma Z$ couplings. These two couplings in the Standard Model (SM) 
are both loop-induced therefore small new physics effects may show up as observable deviations from the SM. 
As one example, the expected deviations in the Inert Triplet Model~\cite{Aus2} are shown in  Fig.\,\ref{fig:1} for the $e^+e^- \to h \gamma$ cross section and the $h \to \gamma \gamma$ branching ratio, which suggests that, depending on model parameters, the deviations can be as large as 100$\%$. 

A usual method to measure the $h\gamma\gamma$ and $h \gamma Z$ couplings is to use branching ratios of $h \to \gamma \gamma / \gamma Z$ decays. It is, however, challenging to measure the  $h \to  \gamma Z$ branching ratio: at the HL-LHC a 5$\sigma$ significance is expected~\cite{Cepeda:2019klc} and at the ILC only a significance of 2.3$\sigma$ is expected~\cite{Aus3}. As a complementary method we study these couplings in a production process at the ILC, $e^+e^- \to h \gamma$. A full simulation analysis in $h\to b\bar{b}$ channel has been reported in~\cite{Aus12}. In this paper we focus on
a new full simulation analysis of the fully hadronic $h\to WW^*$ channel.
  
%, for example some new heavy particles contributing to the loop,
  
This paper is organized as follows. 
In section 2, we explain how to measure $h \gamma Z$ coupling. Section 3 introduces our theoretical framework and experimental method. Our simulation framework is described in section 4. Section 5 gives the combined bound using the measurements of the $h \to \gamma Z$ branching ratio and the $e^+e^- \to h \gamma$ cross section. In section 6, we present a full simulation analysis for the fully hadronic $h\to W W^{*}$ channel. Finally, section 7 summarizes our results and concludes this paper.

        % read manual to see what [ht] means and for other possible 

%\section{Two ways to measure $h \gamma Z$ coupling}
%There are two ways to measure $h \gamma Z$ coupling. The first way is measuring branching ratio of $h \to \gamma Z$. This study was reported in LCWS2018~\cite{Aus3}.  

%The second way is measuring the cross section of the $e^+e^- \to \gamma h$ process. We plan to study 2 higgs decay modes: $h \to b \bar{b}$, and $h \to WW^*$. In this paper, I focus on this method. 

\section{Theoretical Framework and Experimental Method}
We use the effective Lagrangian shown in \Equref{Equ:0} to include new physics contributions to the $e^+ e^- \xrightarrow{} h \gamma $ cross section in a model-independent way,
\begin{eqnarray}
{\cal{L}} _ {h \gamma  } = {\cal{L}} _ { \mathrm { SM } } + \frac { \zeta _ { A Z } } { v } A _ { \mu \nu } Z ^ { \mu \nu } h + \frac { \zeta _ { A } } { 2 v } A _ { \mu \nu } A ^ { \mu \nu } h, 
\label{Equ:0}
\end{eqnarray}
where, in addition to the first term from the SM, $\zeta_{AZ}$ and $\zeta_A$ terms represent, respectively, effective $h \gamma Z $ and $h \gamma \gamma $ couplings from new physics. $A_{\mu \nu}$ and $ Z_{\mu \nu}$ are field strength tensors for the photon and the $Z$ boson, respectively, and $v$ is the vacuum expectation value.

\begin{figure}[ht] 
        
        \centering \includegraphics[width=0.9\columnwidth]{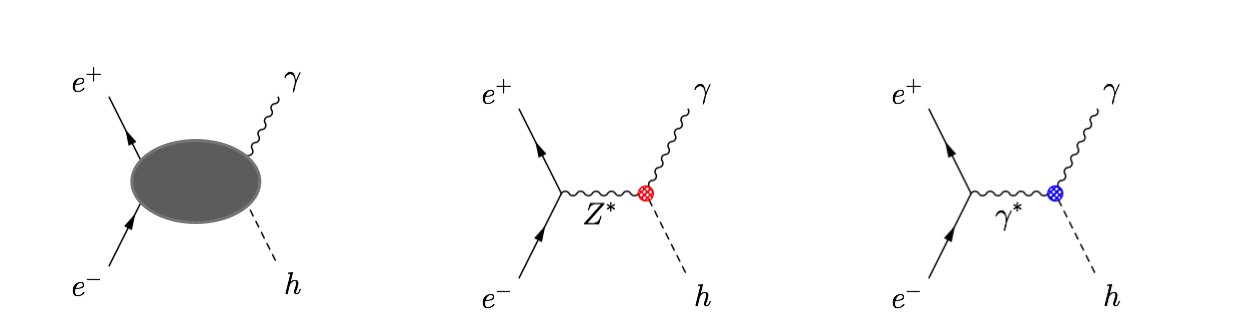}
        \caption{
                \label{fig:5} 
              Diagrams arising from each of the three terms of \Equref{Equ:0}, respectively. }
\end{figure}

\begin{figure}[ht] 
        
        \centering \includegraphics[width=0.9\columnwidth]{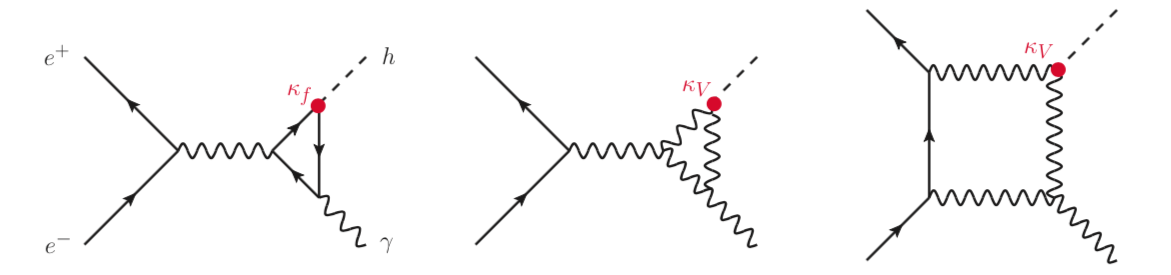}
        \caption{
                \label{fig:7} 
              The loop induced Feynman diagrams in the Standard Model for $e^+e^- \to h \gamma$~\cite{Aus2}}
\end{figure}

\begin{figure}[ht] 
        
        \centering \includegraphics[width=0.6\columnwidth]{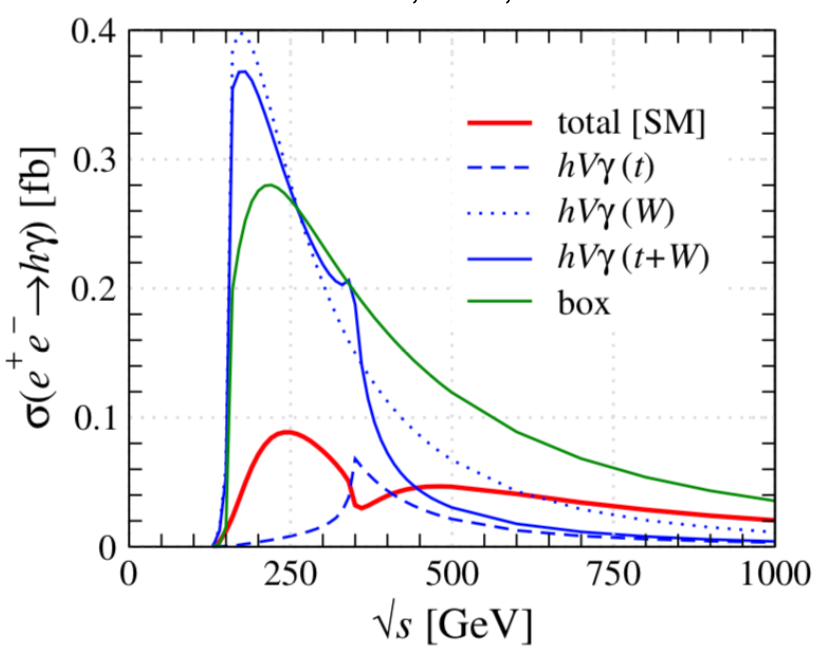}
        \caption{
                \label{fig:20} 
              The contributions from individual diagrams of \Figref{fig:7}. }
\end{figure}

The three terms contribute to the $e^+e^- \to h \gamma$ process via the Feynman diagrams shown in \Figref{fig:5}, where the first SM diagram represents several loop induced diagrams as shown in \Figref{fig:7}. The contributions from individual diagrams of \Figref{fig:7} for unpolarized beams is shown \Figref{fig:20}. We can clearly see that there are significant destructive interferences between these diagrams. 
The SM cross sections at $\sqrt{s}$  = 250 GeV are shown in \Tabref{tbl:1}, which are much less than 1~fb, indicating that experimental measurements would be challenging.
The cross sections including effective $h \gamma Z / h\gamma \gamma$ couplings from new physics, normalized to their SM values, are given 
in \Equref{Equ:1} for beam polarizations $P(e^-,e^+)=(-100\%,+100\%)$ and in \Equref{Equ:2} for $P(e^-,e^+)=(+100\%,-100\%)$, up to interference terms.
\begin{eqnarray}
\frac { \sigma _ { \gamma H } } { \sigma _ { S M } } = 1 - 201 \zeta _ { A } - 273 \zeta _ { A Z }
\label{Equ:1}
\end{eqnarray}

\begin{eqnarray}
\frac { \sigma _ { \gamma H } } { \sigma _ { S M } } = 1 + 492 \zeta _ { A } - 311 \zeta _ { A Z }
\label{Equ:2}
\end{eqnarray}
Since $\zeta_A$ can be constrained by the measurement of the $h \to \gamma \gamma$ branching ratio at the (HL-)LHC, we can extract $\zeta_{AZ}$ 
by measuring the cross section of $e^+e^- \to h \gamma$ at the ILC for just one set of beams polarizations.

\begin{table}[htbp]
\begin{center}
\caption{SM cross sections for different beam polarizations ($\sqrt{s}$  = 250 GeV).}
\label{tbl:1} % spaces are big no-no withing labels
\begin{tabular}{|c|c|c|c|} 
\hline
\multicolumn{1}{|c}{$P_{e^-}$ } & \multicolumn{1}{|c|}{$P_{e^+}$} & \multicolumn{1}{c|}{$\sigma_{SM}$[fb]}  \\
\hline
-100$\%$ & +100$\%$ &  0.35\\
+100$\%$ & -100$\%$	& 0.016\\
-80$\%$ & +30$\%$ & 0.20\\

\hline
\end{tabular}
\end{center}
\end{table}

\clearpage

\section{Simulation Framework}

We use fully-simulated Monte-Carlo (MC) samples produced with the ILD DBD model~\cite{Aus4}. For event generators, we use Physsim~\cite{Aus5} for the signal, and Whizard~\cite{Aus6} for background processes. We include all $e^+e^-\to$ 2-fermion (2f) and 4-fermion (4f) SM processes in the background. ISR and Beamstrahlung effects are included in the event generators. For detector simulation, we use Mokka~\cite{Aus7}, which is based on Geant4~\cite{Aus8}, and for event reconstruction, we use Marlin in iLCSoft~\cite{Aus9}, where particle flow analysis (PFA) is done with PandoraPFA~\cite{Aus10} and flavor tagging is done with LCFI+~\cite{Aus11}.  The analysis is carried out at $\sqrt{s}$=250 GeV, assuming an integrated luminosity of 2 ab$^{-1}$ with $P(e^-,e^+)=(-0.8,+0.3)$.

\section{Combined Result}
Previously, we reported an analysis of the $h \to b \bar{b}$ channel at LCWS2018~\cite{Aus12} that a signal significance of 0.53$\sigma$ is expected for the SM cross section. Using this result and \Equref{Equ:10}, we can set bounds on the parameter $\zeta_{AZ}$,

\begin{eqnarray}
4.1>\frac{\sigma_{\gamma H}}{\sigma_{S M}}=1-201 \zeta_{A}-273 \zeta_{A Z}>0\\
-0.011<\zeta_{AZ}<0.0037,
\label{Equ:10}
\end{eqnarray}
where $\zeta_{A}=0$ is assumed and 4.1 is the 95\% C.L. upper limit estimated with the simplified formula $1.64/significance+1$.
We can set an additional bound in the same way using a previous study~\cite{Aus3} which reported an expected significant of 2.31$\sigma$ for the $h \to \gamma Z$ branching ratio.
\begin{eqnarray}
1.71 > \frac{\mathrm{BR}(h \rightarrow \gamma Z)}{\mathrm{BR}_{S M}}=1+290 \zeta_{A Z} > 0\\
-0.0034<\zeta_{A Z}<0.0024,
\end{eqnarray}
where 1.71 is the 95\% C.L. upper limit. The expected combined 1$\sigma$ bound on $\zeta_{A Z}$ is then 
\begin{eqnarray}
\frac{1}{\sigma_{\Delta \zeta}^{2}}=(290)^{2}(2.31)^{2}+(-273)^{2}(0.53)^{2}\\
-0.0015<\zeta_{A Z}<0.0015.
\end{eqnarray} 

\section{Fully Hadronic $h \to WW^*$ Decay Channel}
\subsection{Event Selection}
The new signal channel study in this paper is $e^+ e^- \to h \gamma $, followed by $h \to WW^*$, where both $W$s decay hadronically. In the final states of the signal events, we expect one isolated monochromatic photon with an energy of $E_{\gamma}=\sqrt{s}/{2}\left( 1- {\left(m_h/\sqrt{s}\right)}^2 \right) = 93$~GeV, where $m_h$ is the Higgs mass. The energy resolution of the electromagnetic calorimeter is typically $\sigma_E=0.16 \times \sqrt{E}$~(GeV), where the photon energy $E$ is in units of GeV~\cite{Aus4}. The energy resolution for the isolated photon is thus around 1.5~GeV.
The main background we expect would be $e^+e^-\to W^+W^-$ with a hard ISR photon.

As pre-selection, we start with identifying one isolated photon with an energy greater than 50 GeV. Sometimes, the reconstruction software PandoraPFA splits calorimetric clusters created by a single high energy photon into several objects. Such split clusters fall within a narrow cone ($\cos\theta_{cone}$=0.998, where $\theta_{cone}$ is cone angle), and are combined into a single photon in the following analysis. The particles other than the photon are clustered into four jets using the Durham algorithm~\cite{Aus13}. A pair of jets among the four jets, which has the invariant 
mass closest to $m_W=80.4$ GeV, is combined to form the on-shell $W$, namely $W_1$.
And the other pair of jets is combined to form the off-shell $W^*$, namely $W_2$. The four jets are combined to form the Higgs boson.

\begin{figure}[ht] 
        
        \includegraphics[width=0.5\columnwidth]{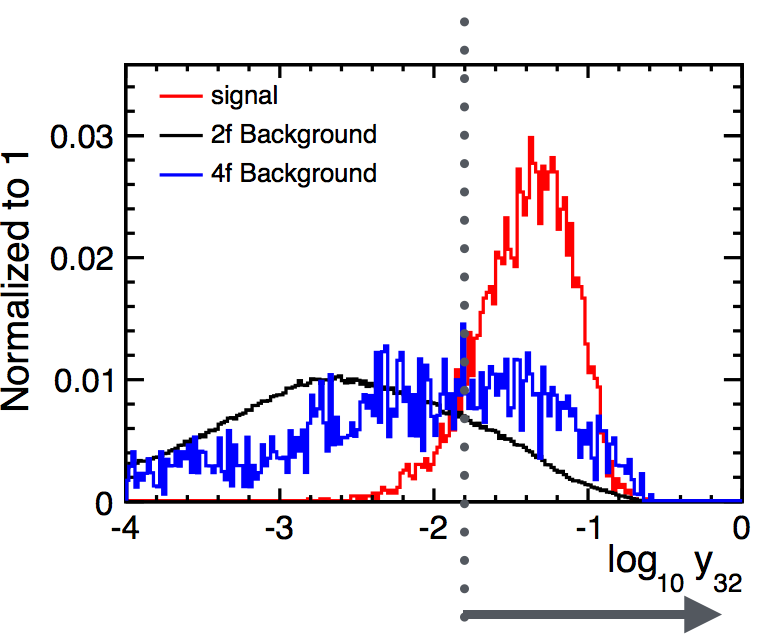}
 \includegraphics[width=0.5\columnwidth]{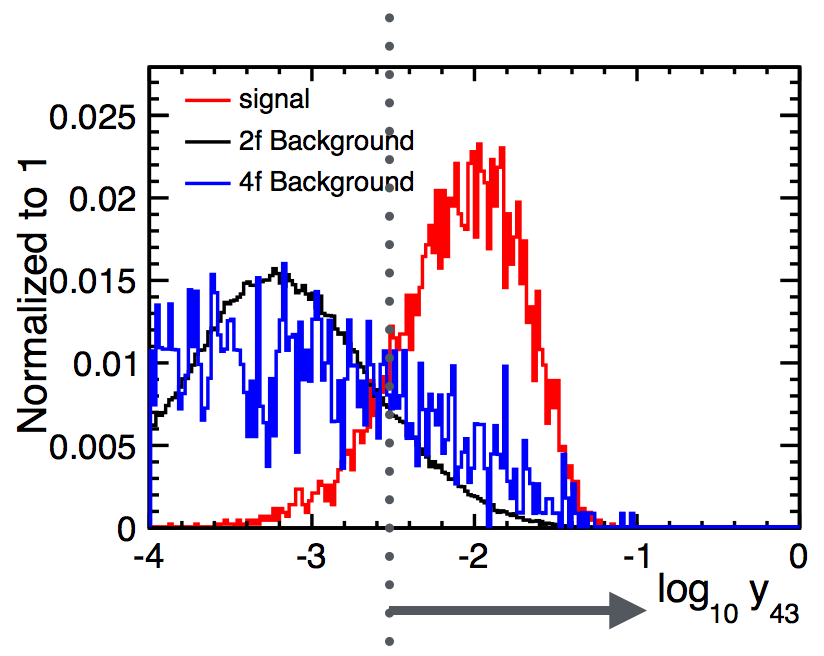}
        \caption{
                \label{fig:8} 
               Distributions of $y_{43}$ and $y_{32}$ for signal and background events. }
               
\end{figure}

%For signal, we applied the cut of the number of decay w to $q\bar{q}$=2.
%caution: this is not a realistic cut, since it is based on MC Truth information which
%will never be known for the real data; it was used here in the simulation as a technical purpose to select the targeted signal events (fully hadronic decay in h->WW*)
As the final selection, we first apply cuts to suppress 2-fermion background. We show the characteristics of signal and 2-fermion background in \Tabref{tbl:bins3}. We require the number of particles in each jet to be greater than 5, and the number of charged particles in each jet greater then 1. Then we demand $\log_{10}(y_{43})>-2.5$ and $\log_{10}(y_{32})>-1.8$, where $y_{mn}$ is the jet distance parameter
defined in the Durham jet-clustering at the step from $m$ jets to $n$ jets. In \Figref{fig:8} the distributions of $y_{32}$ and $y_{43}$ are shown for the signal and 
background events.

\begin{table}[htbp]
\begin{center}
\caption{The characteristics for signal and 2-fermion background events}
\label{tbl:bins3} % spaces are big no-no withing labels
\begin{tabular}{|c|c|c|c|} 
\hline
 \multicolumn{1}{|c|}{Signal} & \multicolumn{1}{c|}{background} & \multicolumn{1}{c|}{Effective cut} \\
\hline
 Many particles in a jet  &  few particles in jet &\\
 & ($e^+e^-/\mu^+\mu^-/\nu\bar{\nu}$) & \# of particles in jet $>$5\\
Many charged particles in a jet & few charged particles in jet 	&  \\
& ($\tau^+\tau^-$)	& \# of charged particles in jet $>$1 \\
large $y_{43}$, $y_{32}$ & relatively small $y_{43}$, $y_{32}$& \\
 & ($q\bar{q}$)	& $y_{43}$, $y_{32}$\\
\hline
\end{tabular}
\end{center}
\end{table}

\begin{figure}[ht] 
        
        \includegraphics[width=0.5\columnwidth]{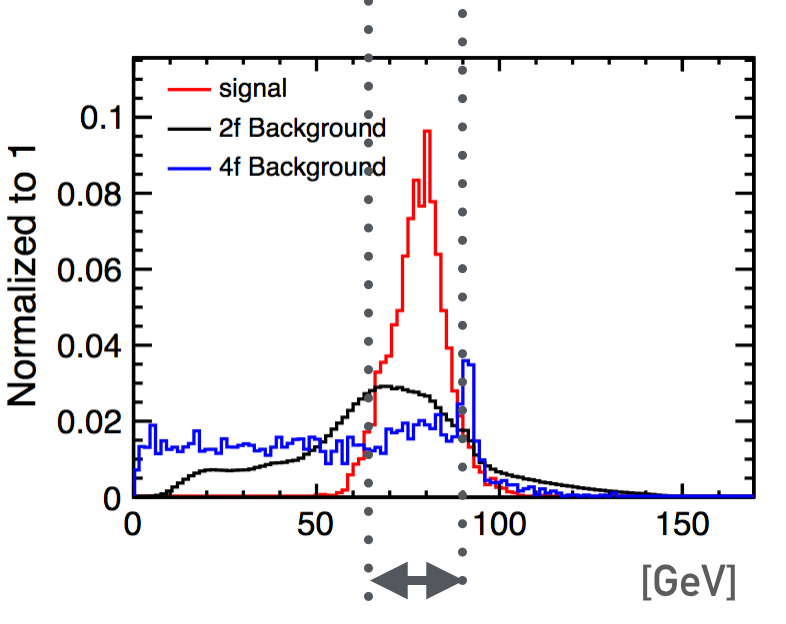}
 \includegraphics[width=0.5\columnwidth]{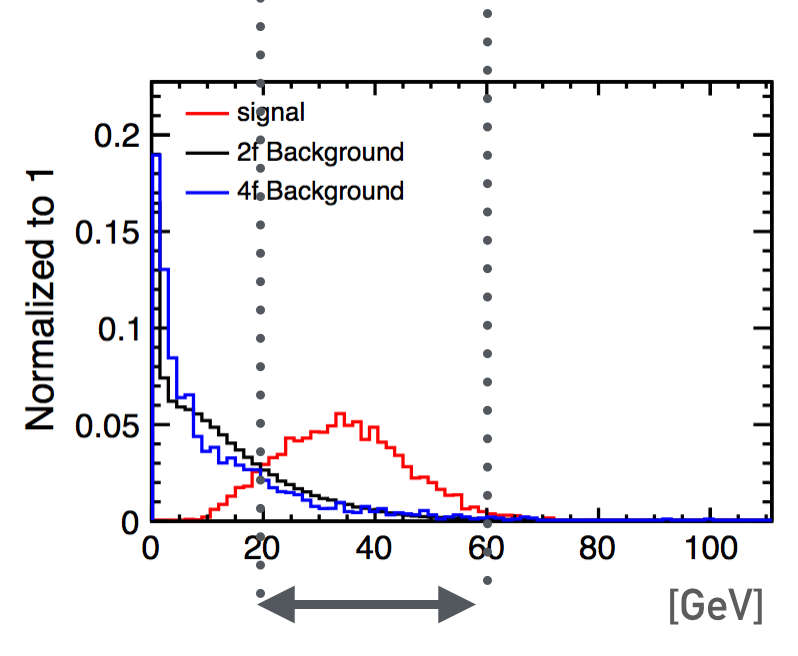}
        \caption{
                \label{fig:15} 
               Distributions of $m_{\rm w1}$ (left) and $m_{\rm w2}$ (right) for signal and background events. }
               
\end{figure}

\begin{figure}[ht] 
        
        \includegraphics[width=0.5\columnwidth]{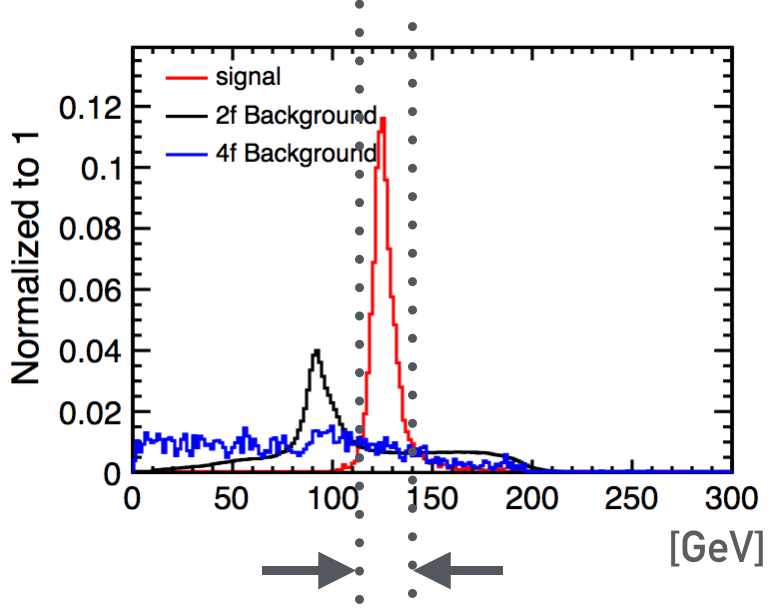}
 \includegraphics[width=0.5\columnwidth]{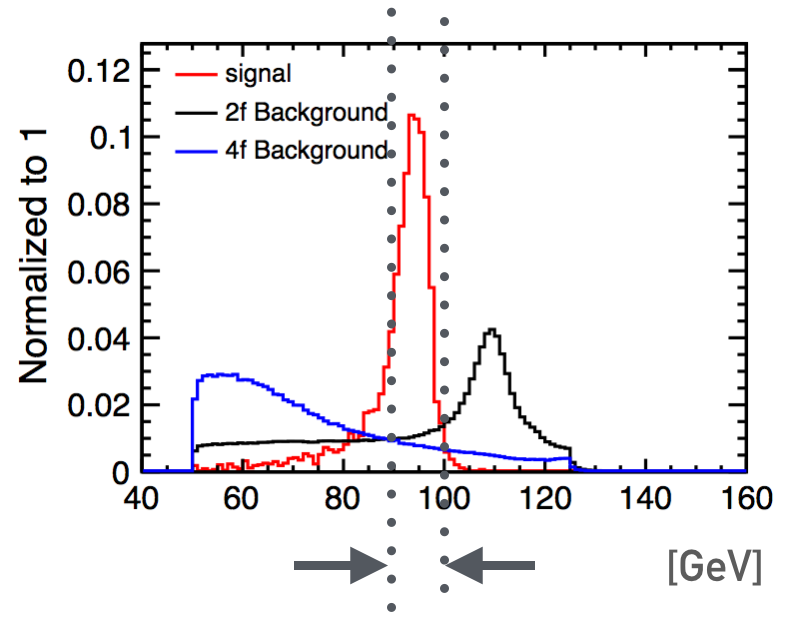}
        \caption{
                \label{fig:14} 
               Distributions of reconstructed Higgs mass (left) and photon energy (right) for signal and background events. }
               
\end{figure}

\begin{figure}[ht] 
        \includegraphics[width=0.5\columnwidth]{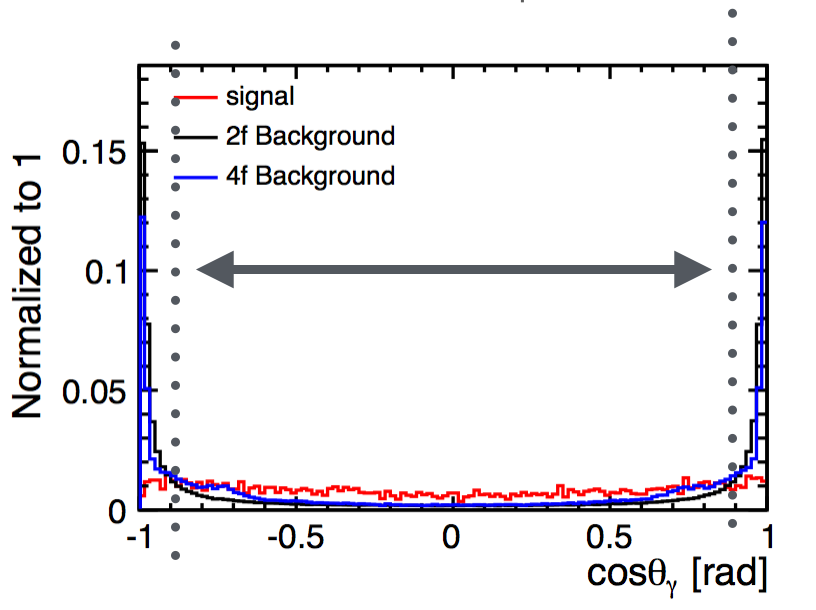}
 \includegraphics[width=0.5\columnwidth]{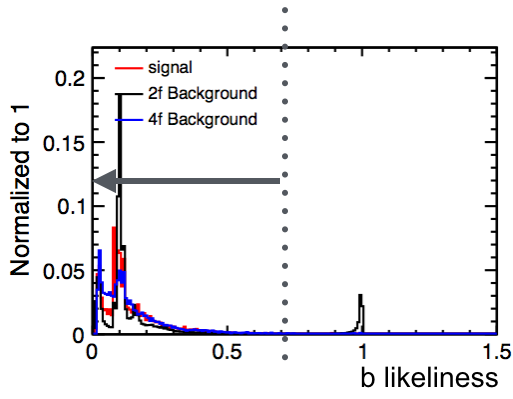}
        \caption{
                \label{fig:12} 
               Distributions of cosine of photon polar angle (left) and the largest  $b$-likeliness for signal and background events.}
\end{figure}

We now try to suppress the 4-fermion background dominated by $e^+e^- \to W^+W^-(\gamma)$. \Tabref{tbl:bins2} shows the characteristics for signal and background events. We apply the following cuts: 65$<m_{\rm w1}<$90, 20$<m_{\rm w2}<$60, 115$<m(4jets)<$135, 90$<E_{\gamma}<$100, where all the numbers are in units of GeV, and $|\cos\theta_\gamma|<0.9$. The distributions of these variables are shown in Figs.\,\ref{fig:15}, \ref{fig:14}, \ref{fig:12}. 

\begin{table}[htbp]
\begin{center}
\caption{Characteristics of signal and 4-fermion background events.}
\label{tbl:bins2} % spaces are big no-no withing labels
\begin{tabular}{|c|c|c|c|} 
\hline
 \multicolumn{1}{|c|}{Signal} & \multicolumn{1}{c|}{background} & \multicolumn{1}{c|}{Effective cut} \\
\hline
 one on-shell $W$  &    no $W$ resonance (such as $ZZ\to 4f$) & $W_1$ mass\\
 one off-shell $W$ & two $W$ resonances ($W^+W^-\to 4f$)	& $W_2$ mass \\
 $h$ resonance  & no $h$ resonance	& Higgs mass\\
monochromatic photon & Energy of photon~small & $\gamma$ energy\\
photon is not forward & photon is very forward & $\gamma$ polar angle \\
\hline
\end{tabular}
\end{center}
\end{table}

\clearpage
As the final cut, we require the largest $b$-likeliness among the four jets (defined as bmax1) to be smaller than 0.7, to suppress events from $e^+e^-\to h\gamma$ followed by $h\to b \bar{b}$ and other 
background events (see \Figref{fig:12}).

The cut values are optimized to maximize the signal signifcance defined as
\begin{eqnarray}
\text {significance} = \frac { N _ { S } } { \sqrt { N _ { S } + N _ { B } } },
\label{Equ:7}
\end{eqnarray}
where $N_S$ and $N_B$ are the numbers of signal and background events, respectively.

\subsection{Result}
\Tabref{tbl:bins} gives the numbers of signal and background events, as well as the signal significance after each cut. The significance is defined by \Equref{Equ:7}. 
After all the cuts, the signal significance is expected found to be 0.09$\sigma$, for the SM signal process $e^+e^-\to h \gamma$ followed by the  fully hadronic  $h\to WW^*$ decay. 
\begin{table}[htbp]
\begin{center}
\caption{Reduction table for the signal and background events after each cut}
\label{tbl:bins} % spaces are big no-no withing labels
\begin{tabular}{|c|c|c|c|} 
\hline
\multicolumn{1}{|c}{ } & \multicolumn{1}{|c|}{Signal} & \multicolumn{1}{c|}{background} & \multicolumn{1}{c|}{Significance} \\
\hline
Expected & 40.2 &   3.14$\times 10^8$ & 0.005\\
Pre selection & 37.7	& 6.10$\times10^7$	&0.01 \\
\# of particle$>$5  & 32.0	& 1.12$\times10^7$	&0.01\\
\# of charged particle $>$1 & 25.9	& 6.65$\times10^6$	&0.01 \\
$\log10(y_{43})$ $>$-2.5 &   	&  	& \\
$\log10(y_{32})$  $>$-1.8 & 20.9 	& 1.52$\times10^6$	&0.02\\
65$<m_{\rm w1}<$90 & 18.9 	&8.60$\times10^5$	&0.02\\
20$<m_{\rm w2}<$60 & 17.4	&5.59$\times10^5$	&0.02\\
115$<m(4jets)<$135 & 15.7 	&7.44$\times10^4$	&0.06\\
90$<E_\gamma<$100 & 11.8 	& 2.73$\times10^4$	&0.07\\
-0.9 $< cos\theta<$0.9 & 10.3 	&1.45$\times10^4$	&0.09\\
b likliness$<$0.7 & 10.0 	&1.36$\times10^4$	&0.09\\
\hline
\end{tabular}
\end{center}
\end{table}

\section{Summary and Conclusions}
In this paper, we have studied measurements of the $h \gamma Z$ coupling in two different ways, the first method is to measure the branching ratio of the $h \to \gamma Z$ decay and the other to measure the cross section for the $e^+e^- \to h \gamma$ process at the 250 GeV ILC, assuming 2 ab$^{-1}$ data collected by the International Large Detector (ILD). We found the expected 1$\sigma$ bound on the effective $h\gamma Z$ coupling ($\zeta_{AZ}$): $-0.0015<\zeta_{AZ}<0.0015$, combining measurements of the cross section for $e^+e^- \to h \gamma$ followed by $h \to b \bar{b}$ and the $h \to \gamma Z$ branching ratio. 
We have also performed a full simulation for the fully hadronic $h \to WW^*$ channel and found the expected signal significance of 0.09$\sigma$ for beam polarizations of $P(e^-,e^+)=(-80\%,+30\%)$.

We are planning to improve our analysis by adding the $h \xrightarrow{} WW^*$ semi-leptonic channel. After the analysis of $h\to WW^*$ channel is completed, we will combine the bounds on $\zeta_{AZ}$ from different channels, and translate the combined bound into that on Dimension-6 operators. We will then investigate the role of the combined bound in one global EFT analysis.

\section*{Acknowledgements}
We would like to thank the LCC generator working group and the ILD software working group for providing the simulation and reconstruction tools and producing the Monte Carlo samples used in this study.
This work has benefited from computing services provided by the ILC Virtual Organization, supported by the national resource providers of the EGI Federation and the Open Science GRID.

\end{document}